\documentclass[a4paper]{jpconf}
\usepackage{graphicx}
\usepackage{amsmath}

\newcommand{\ie}{i.e.,~}
\newcommand{\eg}{e.g.,~}

\newcommand{\bhac}{\texttt{BHAC}~}

\newcommand{\harmthreed}{\texttt{HARM3D}~}
\newcommand{\amrvac}{\texttt{MPI-AMRVAC}~}
\newcommand{\paramesh}{\texttt{PARAMESH}~}
\newcommand{\athenapp}{\texttt{Athena++}~}

\newcommand{\bhoss}{\texttt{BHOSS}~}
\newcommand{\raptor}{\texttt{RAPTOR}~}

\begin{document}
\title{The Black Hole Accretion Code: adaptive mesh refinement and constrained transport}

\author{HR Olivares S\'anchez, O Porth, Y Mizuno}

\address{Institute for Theoretical Physics, Max-von-Laue-Str. 1, 60438 Frankfurt, Germany.}

\ead{olivares@th.physik.uni-frankfurt.de}

\begin{abstract}
With the forthcoming VLBI images of Sgr A* and M87, simulations of accretion
flows onto black holes acquire a special importance to aid with the interpretation of the
observations and to test the predictions of different accretion scenarios, including those
coming from alternative theories of gravity. The Black Hole Accretion Code (\bhac) is a
new multidimensional general-relativistic magnetohydrondynamics (GRMHD) module for the
\amrvac framework. It exploits its adaptive mesh refinement techniques (AMR) to solve
the equations of ideal magnetohydrodynamics in arbitrary curved spacetimes with a significant
speedup and saving in computational cost. In a previous work, this was shown using a Generalized Lagrange Multiplier (GLM) to enforce the solenoidal constraint of the magnetic field. While GLM is fully compatible with \amrvac's AMR infrastructure, we found that  simulations were sensible to the divergence control technique employed, resulting in an improved behavior for those using Constrained Transport (CT). However, cell-centered CT is incompatible with AMR, and several modifications were required to make AMR compatible with staggered CT. We present here preliminary results of these new additions, which achieved machine precision fulfillment of the solenoidal constraint and a significant speedup in a problem close to the intended scientific application.
\end{abstract}

\section{Introduction}

The Black Hole Accretion Code \bhac is an extension of the \amrvac framework
to perform General Relativistic Magnetohydrodynamics (GRMHD) simulations in 1, 2, and 3 dimensions
using finite volume methods and a variety of modern numerical methods,
described more in detail in \cite{Porth2017}.
It exploits \amrvac's infrastructure for parallelization
and block-based automated Adaptive Mesh Refinement (AMR),
resulting in a significant saving in computational time and resources.

In fact, despite the variety of General Relativistic Hydrodynamics and Magnetohydrodynamics
codes currently available
\cite{Hawley84a, Koide00, DeVilliers03a,
  Gammie03, Baiotti04, Duez05MHD0, Anninos05c, Anton05, Mizuno06,
  DelZanna2007, Giacomazzo:2007ti, Radice2012a, Radice2013b,
  McKinney2014, Etienne2015, White2015, Zanotti2015, Meliani2017}
aside from some exceptions as
\cite{Anninos05c,Zanotti2015b,White2015}
AMR is still not a commonly exploited tool.

However, AMR capabilities can be extremely useful for some problems
that are currently computationally prohibitive for most codes.
These include resolving simultaneously the formation and propagation of relativistic jets
from black holes, due to the interaction between very different physical scales
(see \eg \cite{Mizuno2015}), or tilted accretion disks,
where the highly asymmetric evolution prevents the use of the static stretched grid commonly
employed to increase resolution at the equator.

The most immediate application envisaged for \bhac is the simulation
of Sgr A* and M87, the two primary targets of the Event Horizon Telescope (EHT).
Both these objects belong to the class of advection dominated accretion flows (ADAFs),
for which ideal magnetohydrodynamics without radiation feedback constitutes a reasonable approximation
of the plasma properties. In order to properly study the impact of plasma and gravitational conditions
on the EHT images, \bhac is coupled to the General Relativistic Radiative Transfer (GRRT)
codes \bhoss \cite{Younsi2017} and \raptor \cite{Bronzwaer2017}.

An important motivation for our research is the possibility to distinguish departures
from General Relativity in the images obtained by the EHT.
For this reason, \bhac is designed with a modular structure
that can handle arbitrary spacetimes, including numerical ones
as well as those coming from alternative theories of gravity.
For instance, \cite{Mizuno2017} successfully performed GRMHD simulations of accretion
flows onto a dilaton black hole in the Einstein-Maxwell-Axion-Dilaton theory of gravity.
As an example of another application, the code has recently been used to study quasi-periodic-oscillations
(QPOs) in accretion discs around neutron stars \cite{deAvellar2017}.

In a previous work \cite{Porth2017}, we tested \bhac in several standard problems and validated it
by comparing the results of some of them to those obtained in control simulations performed using the
well known code \harmthreed \cite{Gammie03,Noble2009}.

The new additions presented in this work concern changes in the AMR infrastructure, necessary to allow
AMR to operate simultaneously with Constrained Transport (CT),
a divergence control scheme which already showed considerable advantages with respect
to GLM, the technique used in \cite{Porth2017} (see section \ref{sec:divb} for more details).
One of these advantages is the ability to keep a discretization of $\nabla\cdot\mathbf{B}$ equal to zero to machine precision.
In fact, since no cell-centered divergence-free discretization is currently known to be compatible with AMR \cite{Toth2002},
and staggered versions require special divergence-free prolongation and restriction operators
for face- and edge-allocated quantities, GLM was the only such technique available for AMR simulations in \bhac.

The paper is organized as follows: section \ref{sec:grmhd} summarizes the equations and the formulation of GRMHD used in the code;
section \ref{sec:numerics} briefly describes the numerical methods employed in the code focusing on the newly implemented
divergence control strategies; and section \ref{sec:tests} reports on two numerical tests performed.
Throughout this work, we employ geometrized units ($G=c=1$) and use the Einstein summation convention.
Greek indices run from 0 to 3, while Latin indices run from 1 to 3.

\section{Equations of GRMHD}
\label{sec:grmhd}
The equations of ideal GRMHD are those of particle conservation, local conservation of energy-momentum and the homogeneous Maxwell equations

\begin{equation}
\nabla_{\mu} (\rho u^{\mu}) = 0 \,, \
\nabla_{\mu} T^{{\mu \nu}} = 0 \,, \ \text{and} \
\nabla_{\mu}\, ^{*}\!F^{\mu\nu} = 0 \,, \label{eq:grmhd}
\end{equation}

\noindent
where $\nabla_{\mu}$ denotes the covariant derivative, $\rho$ is the particle number in the fluid 
frame, $u^\mu$ the fluid 4-velocity, $T^{{\mu \nu}}$ the energy-momentum 
tensor and $ ^{*}\!F^{\mu\nu}$ the dual of the Faraday tensor $F^{\alpha\beta}$.

The Faraday tensor and its dual are such that, for a frame moving at 4-velocity $n^\nu$,
the electric and magnetic fields are given by

\begin{equation}
E^\mu = F^{\mu\nu}n_\nu \ \ \text{and} \ \ B^\mu = ^{*}\!F^{\mu\nu}n_\nu \,.
\label{eq:Faraday}
\end{equation}

\noindent
In ideal MHD, only the magnetic field is evolved (using the homogeneous Maxwell equations),
since the electric field is determined by the ideal MHD condition, which requires that
the electric field in the frame co-moving with the fluid is
$e^\mu = F^{\mu\nu}u_\nu =0$.

To formulate system (\ref{eq:grmhd}) as a set of evolution equations, we use the 3+1 decomposition of spacetime (see for example \cite{Alcubierre:2008} and \cite{Rezzolla_book:2013}).
The spacetime is sliced into spacelike 3-dimensional hypersurfaces with metric $\gamma_{ij}$.
The 4-dimensional line element is expressed as

\begin{equation}
ds^2 = -\alpha^2 dt^2 + \gamma_{ij}(dx^i+\beta^i dt)(dx^j+\beta^j dt) \, ,
\end{equation}
\noindent
where $\alpha$ and $\beta^i$ are called the lapse function and the shift vector.
The 4-velocity of the Eulerian observers is just the vector normal to each hypersurface,
$n_\mu = -\alpha \nabla_\mu t$.
The 4-metric can then be decomposed as $g_{\mu\nu}=\gamma_{\mu\nu} - n_\mu n_\nu$; therefore, $\gamma^\mu\,_\nu$
acts as a projection operator on the hypersurface.

When projecting the equations of system (\ref{eq:grmhd}) along $n_\mu$ and $\gamma^\mu\,_\nu$,
the result is a set of conservation equations with geometry-dependent sources

\begin{equation}
\partial_t (\sqrt{\gamma} \mathbf{U}) +
\partial_i (\sqrt{\gamma} \mathbf{F}^i) = \sqrt{\gamma} \mathbf{S}
\label{eq:conservation}
\end{equation}
\noindent
and the solenoidal constraint for the magnetic field $\partial_i \sqrt{\gamma} B^i = 0$,
which results from the projection $n_\mu\nabla_\nu\,^{*}\!F^{\mu\nu}$.
Here, $\gamma$ is the metric determinant, and the vectors of conserved quantities $\mathbf{U}$,
fluxes $\mathbf{F}^i$, and sources $\mathbf{S}$ are given by

\begin{align}
\boldsymbol{U} = 
\left[
\begin{array}{c}
D  \\
S_{j}  \\
\tau \\
B^{j}
\end{array}
\right] \,, \ 
\boldsymbol{F}^{i} = 
\left[
\begin{array}{c}
\mathcal{V}^{i} D \\
\alpha W^{i}_{j} - \beta^{i} S_{j} \\
\alpha (S^{i}-v^{i} D) - \beta^{i} \tau \\
\mathcal{V}^{i}B^{j} - B^{i}\mathcal{V}^{j}
\end{array}
\right] \, \text{and} \ 
\boldsymbol{S} = 
\left[
\begin{array}{c}
0  \\
\frac{1}{2}\alpha W^{ik}\partial_{j}\gamma_{ik} + S_{i}\partial_{j}\beta^{i} - U\partial_{j}\alpha \\
\frac{1}{2} W^{ik} \beta^{j} \partial_{j} \gamma_{ik} + W_{i}^{j}\partial_{j}\beta^{i} - S^{j} \partial_{j} \alpha \\ 
0
\end{array}
\right] \,,\label{eq:vectors}
\end{align}

\noindent
where $\mathcal{V}^{i}:=\alpha v^{i} - \beta^{i}$ are the {\it transport velocities},
and the others variables are quantities in the Eulerian frame:
$D=-\rho u^\nu n_\nu$ is the number density, $S_i=n_\mu\gamma_{\nu i} T^{\mu\nu}$ the covariant 3-momentum,
$U=n_\mu n_\nu T^{\mu\nu}$ the total energy and $W^{ij}=\gamma_{\mu i}\gamma_{\nu j}T^{\mu\nu}$ the spatial stress tensor.
Evolving $\tau=U-D$ instead of $U$ makes the evolution more accurate in regions
of low energy and allows to recover the Newtonian limit.

Evolution cannot be carried out using only conservative variables,
since the computation of some quantities in the expressions for the fluxes
requires the knowledge of the primitive variables
$\mathbf{P}=\left[\rho, \Gamma v^i, p, B^i \right]$.
Here, $\Gamma$ is the Lorentz factor,
$v^i=u^i/\Gamma - \beta^i/\alpha$ and $p$ is the pressure in the fluid frame.
While it is straightforward to find $\mathbf{U}(\mathbf{P})$,
$\mathbf{P}(\mathbf{U})$ requires numerical inversion.
To this end, \bhac extends the vector $\mathbf{U}(\mathbf{P})$
by the auxiliary variables $\mathbf{A}=\left[\Gamma, \xi \right]$, where
$\xi:=\Gamma^2\rho h$ and $h$ is the specific enthalpy.
The inversion process then consists of finding $\mathbf{A}$ compatible with
$\mathbf{U}$ and $\mathbf{P}$.

\section{Numerical methods and implementation}
\label{sec:numerics}
In this section we will describe briefly the numerical methods employed in \bhac,
focusing on generalities of the finite volume implementation
and the new features of staggered-mesh based divergence control methods
and adaptive mesh refinement for the staggered variables.

For an in depth description of the methods available in the code, including
equations of state, coordinates and handling of the metric data structure,
reconstruction schemes, Riemann solvers and procedures for
primitive variable recovery, we refer the reader to \cite{Porth2017}.

\subsection{Finite volume scheme}
\label{sec:finite-volume}
To obtain the finite-volume scheme used by \bhac,
we discretize the domain into control volumes $\Delta V_{i,j,k}$
and integrate equation (\ref{eq:conservation}) over each of them.
This leads to the equations of evolution for the average
of the conserved quantities inside each cell,

\begin{align}
\begin{split}
\frac{d\boldsymbol{\bar{U}}_{i,j,k}}{dt} = 
- \frac{1}{\Delta V_{i,j,k}}
\Biggl[
&\boldsymbol{F^{1}}\Delta S^{1}\bigr|_{i+1/2,j,k} - \boldsymbol{F^{1}}\Delta S^{1}\bigr|_{i-1/2,j,k} + \\
&\boldsymbol{F^{2}}\Delta S^{2}\bigr|_{i,j+1/2,k} - \boldsymbol{F^{2}}\Delta S^{2}\bigr|_{i,j-1/2,k} + \\
&\boldsymbol{F^{3}}\Delta S^{3}\bigr|_{i,j,k+1/2} - \boldsymbol{F^{3}}\Delta S^{3}\bigr|_{i,j,k-1/2} \Biggr]+ \boldsymbol{\bar{S}}_{i,j,k} \,.
\label{eq:fvolume}
\end{split}
\end{align}

\noindent
The quantities as $\boldsymbol{F^{1}}\Delta S^{1}\bigr|_{i+1/2,j,k}$ are integrals
of the fluxes over the surfaces $\Delta S^{1}\bigr|_{i+1/2,j,k}$
bounding the control volume and
$\boldsymbol{\bar{S}}_{i,j,k}$ is the volume average of the sources.
Both kinds of integrals are approximated to second order, by assigning to $\boldsymbol{F^{n}}$ ($\boldsymbol{n}=1,2,3$)
the point value of the flux at the interface center and to $\boldsymbol{\bar{S}}_{i,j,k}$
the point value at the cell barycenter.
$\boldsymbol{F^{n}}$ is obtained through
the approximate solution of a Riemann problem at the interface,
and static integrals such as cell volumes, interface areas and barycenter positions are
calculated at initialization using fourth-order Simpson's rule and stored in memory.
Equation \ref{eq:fvolume} can then be solved using the integrators present in the 
\amrvac toolkit. These include the simple predictor-corrector, the third order Runge-Kutta RK3
\cite{Gottlieb98} and the strong-stability preserving $s$-step, $p$th-order RK schemes SSPRK($s$,$p$) schemes: SSPRK(4,3), SSPRK(5,4) due to \cite{Spiteri2002}. (For implementation details, see \cite{Porth2014}.)

\subsection{Divergence control}
\label{sec:divb}
Although the induction equation can also be expressed in the form of equation (\ref{eq:conservation}),
using the finite volume scheme of equation (\ref{eq:fvolume}) alone to evolve the magnetic field
usually results in the creation and rapid growth of numerical magnetic monopoles,
driving the evolution towards flagrantly unphysical states.
In order to keep violations to $\nabla\cdot\mathbf{B}=0$ small,
three schemes are available in \bhac that can be used together with AMR.
The first one is the scheme known as the {\it Generalized Lagrange Multiplier} (GLM)
of the Maxwell equations, a generalization of the Dedner scheme \cite{Dedner:2002}
used in Newtonian MHD. This method consists in solving an additional evolution equation
that has the effect of damping
and advecting away the violations to $\nabla\cdot\mathbf{B}=0$.
GLM has already been applied to GRMHD by \eg \cite{Palenzuela:2008sf}.
Though this technique can be straightforwardly included in \bhac's algorithm,
some of its Newtonian versions have been shown to suffer from spurious oscillations
in the magnetic energy and from an artificial growth of the magnetic fields,
effects attributed, respectively,
to the loss of locality due to the parabolic nature of the additional
equation and to the resulting scheme being non-conservative \cite{ Balsara2004, Mocz2016}.
\begin{figure}[h]
\begin{minipage}{12pc}
\includegraphics[width=14pc]{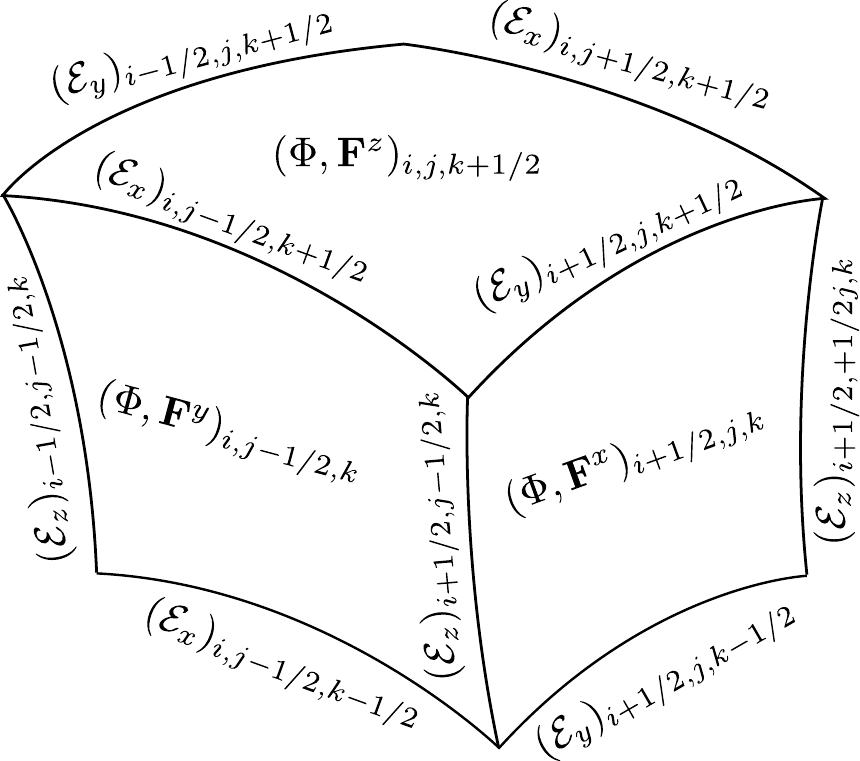}
\end{minipage}\hspace{1.5cm}%
\begin{minipage}{20pc}
\caption{\label{fig:CT_algorithm}
Spatial location of variables for a cell with indices $(i,j,k)$.
Line integrals of the electric field $\mathcal{E}$ are located at its edges, and
magnetic and numerical fluxes $\Phi$ and $\mathbf{F}^i$
(the latter used for the BS algorithm)
are located at its faces.
The rest of variables (not shown) are located at cell centers.}
\end{minipage} 
\end{figure}
The other two available techniques are {\it Constrained Transport} (CT) schemes.
These are obtained by integrating the induction equation
($\gamma^i\,_j\nabla_\mu\,^{*}\!F^{\mu j}=0$) on the {\it boundary} of the control volume,
and their central feature is that they fulfill to machine precision a discretized version of the
solenoidal constraint when constraint-satisfying initial data is supplied
(it can be easily obtained by setting the initial magnetic field as the curl of a potential).
CT was first devised for ideal GRMHD by \cite{Evans1988},
although a similar idea was exploited before in the Yee algorithm \cite{Yee66}.
In these algorithms, the electromagnetic variables are given a special space location (see Figure \ref{fig:CT_algorithm}):
on each face of the cell resides a magnetic flux calculated, \eg as

\begin{equation}
\Phi_{i+1/2,j,k}=\int_{\partial V (x^1_{i+1/2})} \gamma^{1/2} B^1 dx^2dx^3 \,,
\end{equation}

\noindent
and on each edge resides a line integral of the electric field, \eg

\begin{equation}
\label{eq:integralG}
\mathcal{E}_{i+1/2,j+1/2,k} = - \int_{x^3_{k-1/2}}^{x^3_{k+1/2}} \left. E_3\right|_{x^1_{i+1/2},x^2_{j+1/2}} dx^3 \,.
\end{equation}

\noindent
The magnetic flux at each face is updated using the integral form of Faraday's law:

\begin{equation}
\label{eq:CT_update}
\frac{d}{dt}\Phi_{i+1/2,j,k} = \mathcal{E}_{i+1/2,j+1/2,k} - \mathcal{E}_{i+1/2,j-1/2,k} - \mathcal{E}_{i+1/2,j,k+1/2} + \mathcal{E}_{i+1/2,j,k-1/2} \,.
\end{equation}

\noindent
Since each of the line integrals is shared by two faces,
but appears with opposite sign in the time update formula for each of them,
the rate of change of $(\nabla\cdot\mathbf{B})$,
\ie the sum of the rate of change of the outgoing flux through all faces, vanishes.
So far, equation (\ref{eq:CT_update}) is exact.
Each variant of CT arises from different ways of approximating the line integrals
of the electric field.

The two variants available in our code are the method of Balsara \& Spicer (BS)
\cite{Balsara99}
and {\it Upwind Constrained Transport} (UCT) \cite{londrillo2004divergence}.
In BS, $\mathcal{E}_{i+1/2,j+1/2,k}$ is calculated simply as the arithmetic average of the
fluxes obtained by the Riemann solver that correspond to the electric component $E_3$
at the faces $\Delta S_{i+1/2,j,k}$, $\Delta S_{i+1/2,j+1,k}$,
$\Delta S_{i,j+1/2,k}$ and $\Delta S_{i+1,j+1/2,k}$.

A cell-centered version of the BS algorithm,
known as {\it Flux-interpolated Constrained Transport} (flux-CT), was found in \cite{Toth2000}
and is widely used in the literature (see \eg \cite{Gammie03,Noble2009}).
Unlike the staggered version, the cell-centered scheme is not compatible with AMR;
however, it has been reported that in otherwise identical GRMHD simulations performed using GLM and
flux-CT, the latter produced less spurious structures in the magnetic field,
and was able to preserve for a longer time an exact stationary solution \cite{Porth2017}.
This provided a strong motivation for us to implement the staggered algorithm in \bhac
to gain the advantages of AMR.
Nevertheless, BS has also known deficiencies \cite{Flock2010} that can be overcome
by upwinding the electromotive force,
as it is done in the the algorithm by Gardiner \& Stone \cite{Gardiner2005} and in UCT.
The latter, also implemented in \bhac, is another staggered algorithm, devised to incorporate the
correct continuity and upwind properties
of the magnetic field by using limited reconstructions and by taking into account
the transport velocities. In contrast to BS, UCT has the additional property that
it reduces to the correct 1-dimensional limit when the correspondent symmetry is assumed.
For details on the specific UCT implementation in \bhac,
we refer the reader to a more complete work \cite{Olivares2017}, currently in preparation.

\subsection{Adaptive mesh refinement}
\label{sec:amr}
Most of the infrastructure for AMR is inherited from the \amrvac toolkit.
The grid is a fully adaptive block-based octree (in 3D) with a fixed refinement factor of two
between successive levels.
Operations on the grid as time update, IO and problem initialization are performed on
a loop over a Morton Z-order curve.
The time step is calculated globally and is the same for all levels,
thus load-balancing is simply done by cutting the space-filling curve in equal parts
and distributing them among the MPI-processes.
This strategy is applied in various astrophysical codes, for
example in those employing the \paramesh library
\cite{MacNeice00,Fryxell2000,Zhang2006}
, or the recent \athenapp
framework \cite{White2016}.
Refinement can be triggered in a completely automated way
either using the L\"ohner scheme \cite{Loehner87} or user defined prescriptions.
Details on the prolongation and restriction operations and the ghost cell exchange
can be found in \cite{Keppens2012}.
To ensure machine precision conservation of $\boldsymbol{U}$,
re-fluxing is performed every (partial) time step, \ie the fluxes on the coarse side of coarse/fine interfaces are replaced by the sum
of the co-spatial fluxes on the fine side.

New additions specific to \bhac are divergence-free restriction and prolongation operators
for the staggered variables and an electric field fixing step to avoid producing
numerical monopoles across resolution jumps, and which also consists on replacing the
electric fields on the coarse side with their co-spatial fine representation.
Details about the prolongation operator and the electric field fixing formulas
will be documented in a forthcoming work \cite{Olivares2017}.

\section{Numerical tests}
\label{sec:tests}
\subsection{Validation of the code}
\bhac has been thoroughly validated using the GLM
and the flux-CT schemes for the magnetic field evolution.
This was done by performing several test problems in 1, 2 and 3 dimensions,
as well as comparisons with simulations performed using the code \harmthreed,
as is reported in \cite{Porth2017}.
The results obtained using the newly available features here described
were verified against the validated results to ensure that the implementation was correct.
This will be documented in detail in \cite{Olivares2017}.
In the next sections, we will describe the results of applying AMR
and the staggered BS algorithm in two test problems.

\subsection{Relativistic Orszag-Tang vortex}
The Orszag-Tang vortex \cite{Orszag1979} is a common setup to highlight
the impact of the violations to the solenoidal constraint in the numerical
solution.
Starting from a configuration in which $\nabla\cdot\mathbf{B}=0$
to machine precision, the problem quickly develops magnetic shocks and turbulence,
both challenging conditions for the preservation of the constraint.
In this 2-dimensional, special-relativistic realization of the test,
we set
$\rho=1$, $p=10$, $v^x=\frac{0.99}{\sqrt{2}}\sin y$, $v^y=\frac{0.99}{\sqrt{2}}\sin x$, 
$B^x=-\sin y$ and $B^y=\sin 2x$.
The equation of state is that of an ideal fluid with $\gamma=4/3$.
The domain is the square $x,y \in [0,2\pi]$ with periodic boundary conditions.
We adopt three AMR levels, where the lowest resolution is equivalent to resolve the whole domain
with $64 \times 64$ cells.
The numerical methods to evolve the system are an RK3 integrator with HLL fluxes and the
Koren reconstruction (third order accuracy in smooth parts of the solution \cite{Koren1993}).
The divergence control method is the staggered CT algorithm with arithmetic averaging
and the CFL factor is set to 0.4.

\begin{figure}[h]
\begin{minipage}{26pc}
\includegraphics[width=26pc]{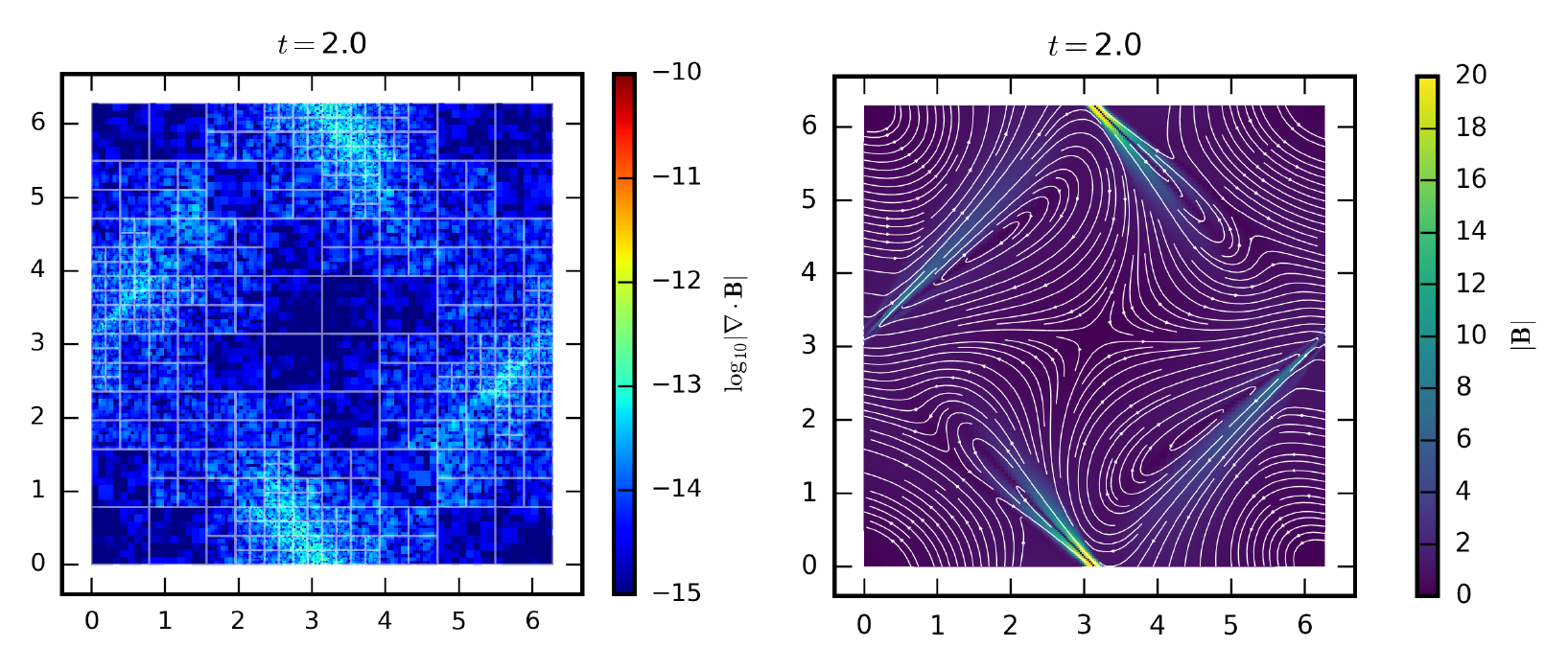}
\end{minipage} 
\hspace{1.5pc}
\begin{minipage}{10pc}
\caption{\label{fig:OT-snapshots}
Snapshots of the relativistic Orszag-Tang problem at $t=2$. \\
{\it Left:} Divergence of the magnetic field and AMR blocks. \\
{\it Right:} Magnetic field intensity and magnetic field lines.}
\end{minipage} 
\end{figure}

Figure \ref{fig:OT-snapshots} shows two snapshots of the evolution.
On the left panel is the divergence of the magnetic field at $t=2$, near the time when the strongest
shocks form.
At that moment, all the three AMR levels are present in the simulation,
and it can be seen that the algorithm is able to keep the largest violations to the level of $10^{-13}$,
in contrast to the $\sim 10^{-1}-10^{0}$ that are produced in a similar set up
with GLM (see \cite{Porth2017}).
The right panel displays the magnetic field intensity and the magnetic field lines at the same time of the simulation,
showing the formation of current sheets
at the same location of the maximum creation of divergence.

\begin{figure}[h]
\begin{minipage}{22pc}
\includegraphics[width=22pc]{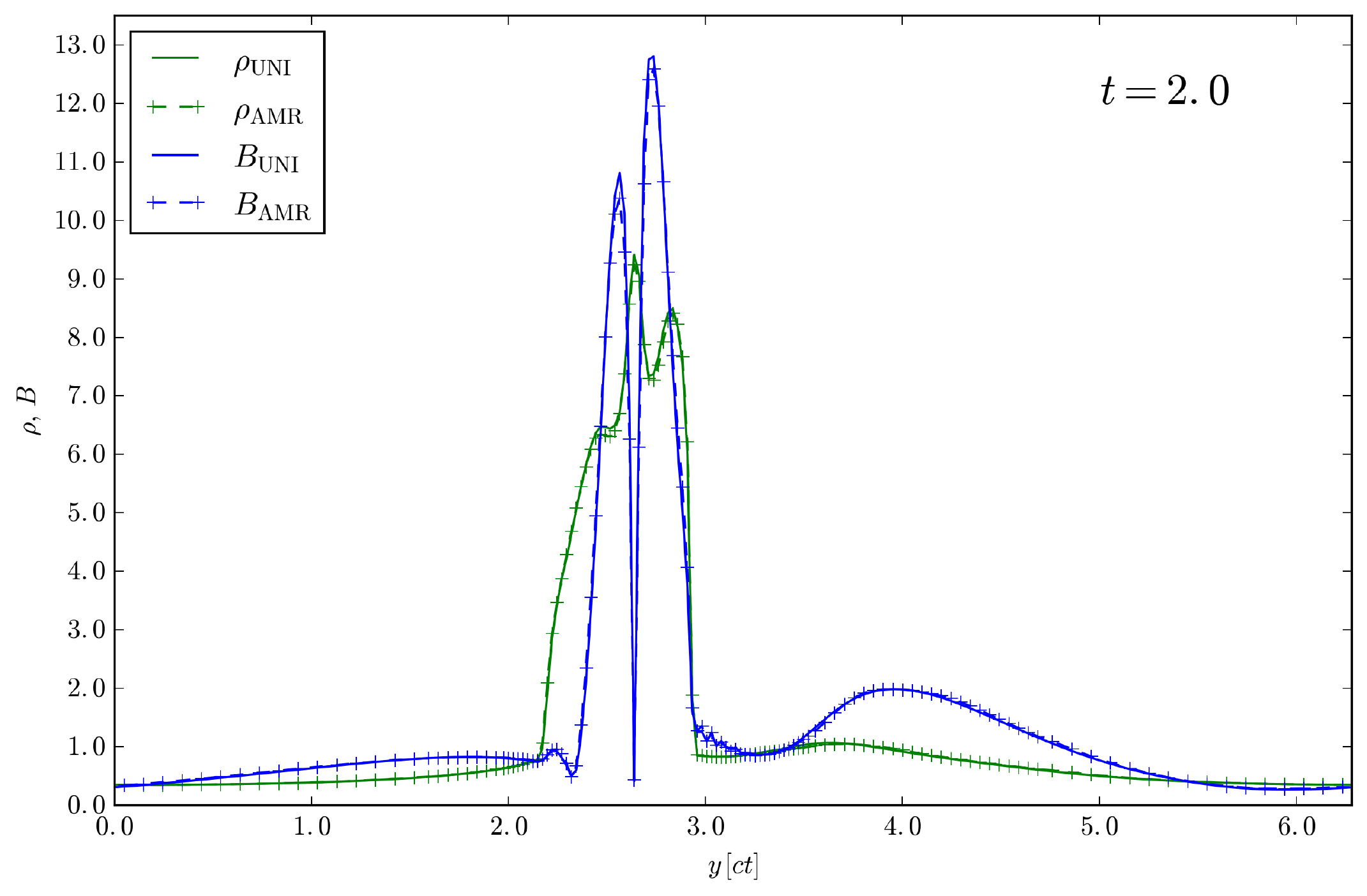}
\end{minipage}\hspace{2pc}%
\begin{minipage}{14pc}
\caption{\label{fig:OT-profile}
Density and magnetic field strength profiles at $y=0.5$ and $t=2$
for uniform resolution and 3-level AMR. The AMR structure is represented by the symbols `$+$'.}
\end{minipage} 
\end{figure}

In order to perform a more quantitative comparison of how well AMR performs, Figure \ref{fig:OT-profile} displays the density and magnetic field strength profiles along a cut
at $y=0.5$ and at the same time, for the simulation described above and for another one
identical except for being run at a uniform high resolution correspondent to that of the highest
AMR level. A very good agreement can be observed between both simulations.
When evolving up to $t=10$ the AMR case obtained a modest speedup factor of 1.35.
This is due to the fact that at later times most of the domain shows large variations
in the quantities that trigger refinement, thus most of the simulation reaches the highest AMR level.

This is, however, not expected to be the case in the intended astrophysical
applications of the code, where likely large parts of the domain are emulating vacuum,
as will be seen in the next section.

\subsection{Magnetized accretion onto Kerr black hole}

\begin{figure}[h]
\begin{minipage}{14pc}
\includegraphics[width=14pc]{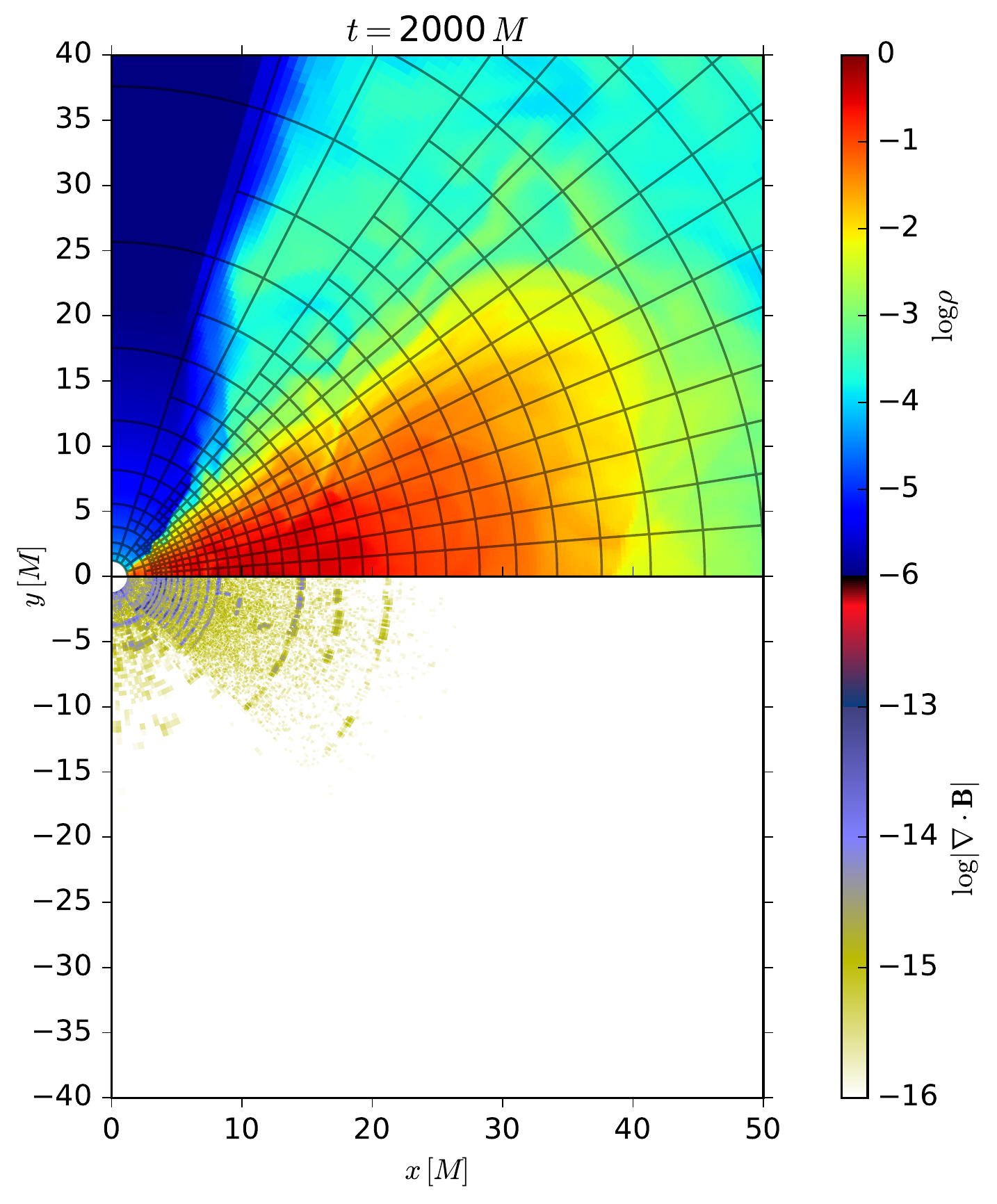}
\caption{\label{fig:amr} {\it Upper half}:  Logarithmic density and AMR blocks at time $t=2000\,M$.}
{\it Lower half}: Divergence of the magnetic field.
\end{minipage}\hspace{2pc}%
\begin{minipage}{22pc}
\includegraphics[width=22pc]{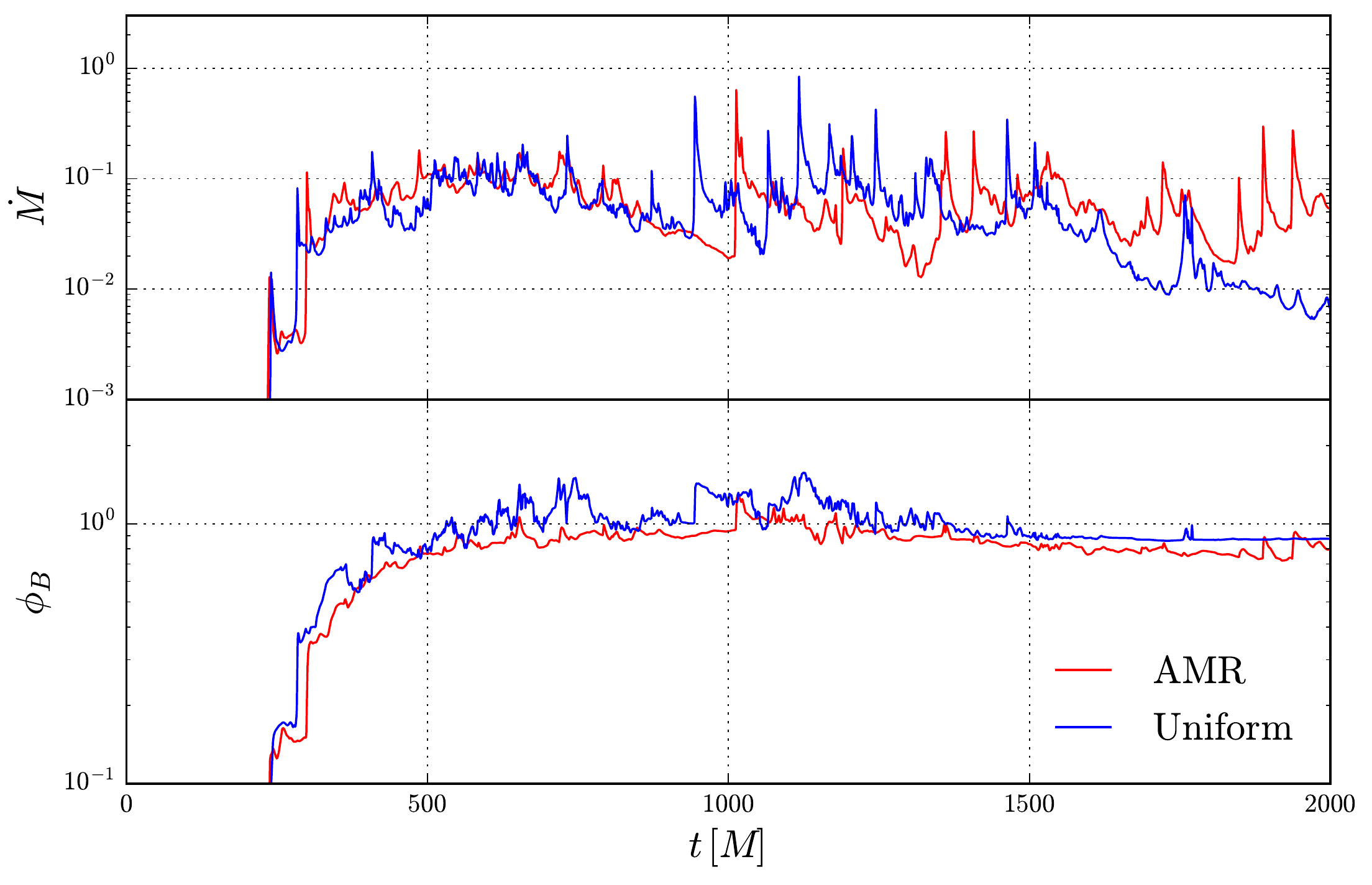}
\caption{\label{fig:acc_rates} Mass accretion rate and absolute magnetic flux through
the horizon for uniform resolution and AMR}
\end{minipage} 
\end{figure}

To study the speedup and accuracy achievable using AMR
in a case closer to the intended application of the code,
we perform two simulations of the same 2D problem of accretion from a magnetized torus onto a Kerr black hole,
one using a grid with uniformly high resolution
and the other one using AMR, with the maximum resolution corresponding to that
of the first simulation.

The spacetime is described using logarithmic Kerr-Schild coordinates,
correspondent to the standard Kerr-Schild coordinates
$r\in [1.213, 2500\,M]$ and $\theta\in [0,\pi]$.
This allows the propagation of the jet over a long distance and prevents signals from the
boundaries to affect the inner region when evolving until $t=5000\,M$.
The spin parameter of the black hole is $a=0.9375$, and the event horizon is located at $r=1.348\,M$.

The fluid obeys an ideal equation of state with $\gamma=4/3$.
As initial condition, we set up an equilibrium torus with inner radius at $r_\mathrm{in}=6\,M$,
and density maximum at $r_\mathrm{max}=12\,M$ (orbital period of $247\,M$ at the density maximum).
A single-loop poloidal magnetic field is built from the vector potential
$A_{\phi} \propto {\rm max} (\rho/\rho_{\rm max} - 0.2, 0)$ and is normalized in such a way
that the minimum plasma $\beta=p_\mathrm{fluid}/p_\mathrm{mag}=100$. To avoid vacuum regions, the rest of the simulation
is filled with a tenuous atmosphere with density $\rho_{\mathrm{fl}} = 10^{-4}
r^{-3/2}$ and fluid pressure $p_{\mathrm{fl}} =
1/3\times10^{-6}\ r^{-5/2}$.
We reset the density or the pressure whenever they fall below these floor values.
To perturb this equilibrium state, random perturbations of $4\%$ are added to the pressure.
This eventually triggers the magneto-rotational instability,
allowing the plasma to accrete.

The simulations are evolved using a two-step predictor-corrector method, TVDLF fluxes,
and PPM reconstruction. The CFL number is set to $0.35$.
For evolving the magnetic field we use staggered CT with arithmetic averaging.
The boundary conditions at the inner and outer radial boundaries are set to zero gradient
in the primitive variables, \ie their values at the last cell of the physical domain
are copied to fill the ghost cells, except for the ingoing component of the velocity,
which is set to zero.
The numerical fluxes and line integrals of the azimuthal electric field
in contact with the polar axis are also set to zero,
since they correspond to integrals over zero-area surfaces and zero-length paths.
The number of cells in the
simulation with uniform high resolution is
$N_r\times N_\theta=800\times 400$.
The simulation with AMR has three levels, the highest with the same resolution as the 
one with the uniform mesh. AMR is triggered automatically using the L\"ohner scheme.

Figure \ref{fig:amr} displays the grid blocks of the AMR simulation at time $t=2000\,M$,
showing how resolution increases in regions with large variations in density.
To perform a more quantitative comparison,
Figure \ref{fig:acc_rates} shows that the mass accretion rate and absolute magnetic flux through 
the horizon for both simulations have comparable magnitude and variability.
While the simulation at uniform resolution required 2324 cpu-hours to reach $t=5000\,M$,
the simulation using three AMR levels required only 327, yielding a significant speedup factor of 7.1.

\section{Conclusion}
\label{sec:conclusion}
\bhac is a new versatile tool to study magneto-hydrodynamic flows in arbitrary
spacetimes in General Relativity and other metric theories of gravity,
which incorporates modern numerical methods and an efficient AMR infrastructure
inherited from \amrvac.

We have made new additions to this infrastructure in order to allow 
staggered-mesh constrained transport
to run with the advantages of AMR.

In this work, we have presented the first results of simulations performed with \bhac using AMR and a CT
together. These indicate that the code is now capable of evolving
efficiently the GRMHD equations in multi-scale problems,
with the solenoidal constraint fulfilled to machine precision.

As a matter of fact, for a problem close to its intended application,
\bhac was able to attain a very significant speedup of 7.1,
which can be of crucial importance, also for \eg performing parameter studies to contrast with
the EHT data.

In a forthcoming work, we will describe in greater detail such modifications as well as other
additions to the numerical methods besides those described in \cite{Porth2017}.

\ack
We would like to express our gratitude to Luciano Rezzolla, Elias Most,
Christian Fromm, Ziri Younsi, Alejandro Cruz Osorio, David Kling, Jonas
K\"ohler and Mariafelicia de Laurentis for useful discussions.
This research is supported by the ERC synergy
grant "BlackHoleCam: Imaging the Event Horizon of Black Holes" (Grant
No. 610058), by ``NewCompStar'', COST Action MP1304, by the LOEWE-Program
in HIC for FAIR, and by the European Union's Horizon 2020 Research and
Innovation Programme (Grant 671698) (call FETHPC-1-2014, project
ExaHyPE). HO is supported in part by a CONACYT-DAAD scholarship. The
simulations were performed on LOEWE at the CSC-Frankfurt and Iboga at ITP
Frankfurt. We acknowledge technical support from Thomas Coelho.

\section*{References}

\bibliography{aeireferences}


\end{document}